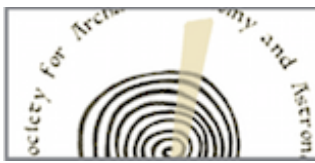

**Peer Reviewed**

**Title:**
Ojibwe Giizhiig Anung Masinaaigan and D(L)akota Makoče Wičaŋȟpi Wowapi: Revitalization of Native American Star Knowledge, A Community Effort

**Author:**
Lee, Annette

**Abstract:**
The Native Skywatchers research and programming initiative focuses on the revitalization of native star knowledge of the Ojibwe and Dakota people. Activities include interviewing elders, culture and language teachers, and creating programming around traditional native star knowledge interlaced with Western science. Star maps, curriculum, hands-on workshops, planetarium shows, and artwork have been designed and delivered. Developed for native and non-native communities in light of the new Minnesota State Science Standards implemented in 2009, presented here are two native star maps that were created by the Native Skywatchers initiative: the Ojibwe Giizhig Anung Masinaaigan (or the Ojibwe Sky Star Map); and the D(L)akota Makoče Wičaŋȟpi Wowapi or (D(L)akota Star Map). This interdisciplinary project includes professional astronomers, professional artists, language and cultural experts, educators, community members and elders.

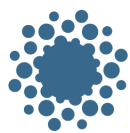

# Ojibwe Giizhiig Anung Masinaaigan and D(L)akota Makoċe Wiċaŋḣpi Wowapi: Revitalization of Native American Star Knowledge, A Community Effort

## Annette S. Lee
St. Cloud State University

**Abstract:**
*The Native Skywatchers research and programming initiative focuses on the revitalization of native star knowledge of the Ojibwe and Dakota people. Activities include interviewing elders, culture and language teachers, and creating programming around traditional native star knowledge interlaced with Western science. Star maps, curriculum, hands-on workshops, planetarium shows, and artwork have been designed and delivered. Developed for native and non-native communities in light of the new Minnesota State Science Standards implemented in 2009, presented here are two native star maps that were created by the Native Skywatchers initiative: the Ojibwe Giizhig Anung Masinaaigan (or the Ojibwe Sky Star Map); and the D(L)akota Makoċe Wiċaŋḣpi Wowapi or (D(L)akota Star Map). This interdisciplinary project includes professional astronomers, professional artists, language and cultural experts, educators, community members and elders.*

*Understanding that men and women throughout history of all cultures, including Minnesota American Indian tribes and communities, have been involved in engineering design and scientific inquiry… For example, Ojibwe and Dakota knowledge and use of patterns in the stars to predict and plan…* (Minnesota Department of Education, Benchmark 3.1.3.2.1, 2009)

## Introduction

Over the past six years the *Native Skywatchers* initiative has addressed the crisis of the loss of the Ojibwe and D(L)akota star knowledge, among the indigenous peoples of Minnesota. There is an urgency to this project because elders are passing, some simply 'weren't listening' when the star stories were being told and at the same time there are many layers of social upheaval on some reservations, including unemployment, addiction, suicides, gangs and lack of clean drinking water. This research and programming is dedicated to rebuilding and reclaiming the native star knowledge, documenting it, disseminating it and developing it. The ideal outcome is that more native people have a meaningful connection to the stars. Through this connection to the stars a sense of cultural pride, a sense of connectedness and purpose is nurtured. Inherently interdisciplinary, our work includes astronomy, culture, language, art, science education, history, and community wellness. The implications of this foundational work are many, including encouraging more native young people to graduate from high school and possibly choose a career in STEM (Science, Technology, Engineering and Math). Currently Minnesota has some of the lowest graduation rates and highest achievement gaps for Native Students in the United States (Matos 2015; Post 2015).

*Team-members*
The *Native Skywatchers* research and programming initiative was founded by Annette S. Lee (mixed-race Dakota Sioux astronomer and artist) in 2007. Other team-members include: Carl Gawboy (Ojibwe), Jeff Tibbetts (Ojibwe), William Wilson (Ojibwe), Jim Rock (Dakota), and Charlene O'Rourke (Lakota). Acknowledgement goes out to two elders who were also part of the team and who have since passed away: Paul Schultz (Ojibwe) and Albert White Hat Sr. (Lakota). What is essential here is that the team is composed of persons of different expertise, scientists, artists, educators, writers, and historians. This is a native-led initiative. The work is collaborative. Due to the history of colonization, assimilation, reservations, etc., much has been lost. No one person holds all the details of the native Ojibwe and D(L)akota star knowledge. Many voices are needed.

*Land and Language*
*Native Skywatchers* is based in Minnesota, which contains eleven Native American reservations: four Dakota (Sioux) and seven Ojibwe (Chippewa). The region also consists of Ojibwe and D(L)akota communities in the surrounding geographical areas of northern Midwestern United States (i.e. South Dakota, North Dakota, Michigan) and southern Canada.

Figure 1. Location of Minnesota in the USA. https://commons.wikimedia.org/wiki/File:Map_of_USA_MN.svg

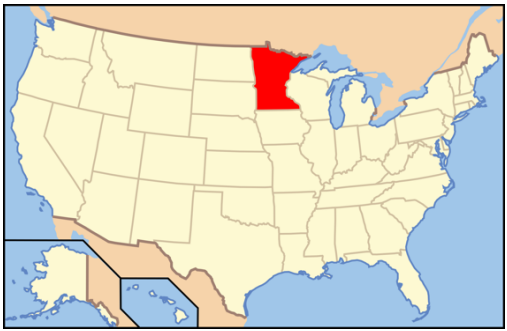

Language is an important part of native star knowledge. Similar to star knowledge, in language there are layers of meaning contained in the Ojibwe and Dakota words. Unfortunately, a person needs to know the language well in order to know

Figure 2. Eleven tribes of Minnesota
http://kspamericanindianproject.wikispaces.com

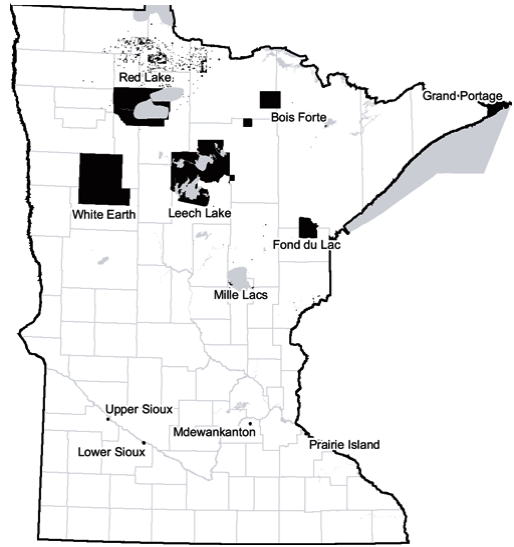

the inherent meaning of the source syllables. For example, the word 'Minnesota' comes from Dakota and is sometimes translated as '*Mni Sota Makoče*' meaning 'land where the waters are so clear they reflect the clouds" (Westerman 2012). One understanding of this is a literal reference to the state containing over 10,000 lakes (10 acres in size or larger). Another interpretation is 'where the waters are so clear they reflect the sky' (Rock 2012). This refers to a pairing of the Milky Way and the Mississippi River, and the teaching '*as it is above; it is below.…*'

The largest river in the area, the Mississippi River, flows generally from north to south. The name "Mississippi" is a mistranslation of the Ojibwe word *Misi-ziibi*, or 'Great/Big River'. (Baraga 1992) The largest city and capitol of Minnesota is Minneapolis-St. Paul, which is located at the confluence of the Mississippi River and the Minnesota River. In Dakota this is a very sacred area containing, *Bdote*, 'the confluence,' and *Wakaŋ Tipi*, 'the sacred cave', which is the Dakota genesis place (Westerman 2012). Here the term 'Dakota' is the abbreviated name of the group of native peoples called *Očeti Šakowiŋ Oyate*, 'Seven Sacred Council Fires.' This council is made up of: four Dakota bands, two Nakota bands and one Lakota band. The Dakota tribe is also known as 'Sioux,' which is a corruption of the Odawa word *Nadouessioux*, meaning 'speakers of a foreign language' (Johnson, 2000). The word *Ojibwe* means 'people who cook outside,' for example roasting rabbit on a fire. (Wilson 2012) Sometimes *Anishinaabe*, 'the people,' is also used for Ojibwe. The corrupted word for the Ojibwe tribe is 'Chippewa.'

Figure 3. Rivers and Streams of Minnesota, Mississippi River seen near center.

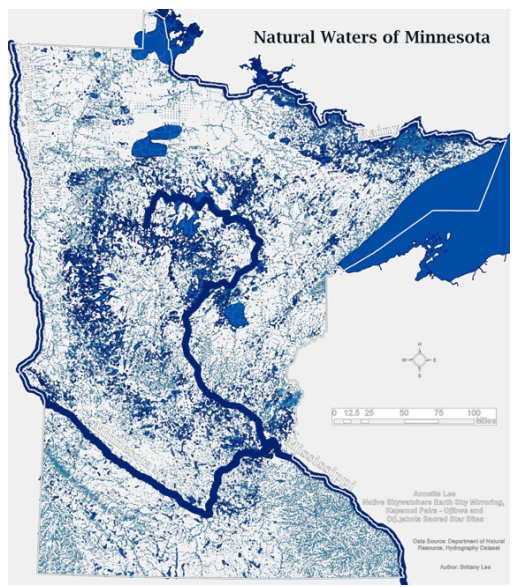

**Resources**

Existing Materials

Prior to 2012 there were exactly two published books dedicated to Ojibwe and D(L)akota sky wisdom: *Talking Rocks: Geology and 10,000 Years of Native American Tradition in the Lake Superior Region* by Morton and Gawboy (2000) and *Lakota Star Knowledge: Studies in Lakota Stellar Theology* by Goodman

(1992). The lack of available resources, especially in light of the newly approved state standards, was one of the first areas for *Native Skywatchers* to address with urgency.

Newly Created Resources

*Star maps*

At the foundation of the current work are two native star maps that were created in 2012: *Ojibwe Giizhig Anung Masinaaigan* and *D(L)akota Makoče Wičaŋḣpi Wowapi*

The Ojibwe Sky Star Map, *Ojibwe Giizhig Anung Masinaaigan*, was painted by the author and William Wilson, with Wilson serving as the language expert. The map was based on the unpublished work of Carl Gawboy. Since the 1960s Gawboy had been interviewing elders and researching Ojibwe star knowledge. Gawboy (2005) was the first to identify the pictographs at Lake Hegman, Boundary Waters as Ojibwe constellations.

The Dakota Star Map, *D(L)akota Makoče Wičaŋḣpi Wowapi*, was painted by the author. The language expert here was Jim Rock. The map was based on the published work of Ron Goodman. In the 1980s Goodman interviewed many Lakota elders from the South Dakota area and published his results in the book *Lakota Star Knowledge* (1992). Several of the elders quoted in Goodman's book are part of the *Native Skywatchers* project.

Both maps are astronomically accurate and visual works of art created to communicate an indigenous perspective of the night sky. Located at the center of both maps is the north celestial pole (NCP) and the North Star (Polaris), which is *Giiwedin Anung* (Ojibwe) and *Wičaḣpi Owaŋjila* (Dakota). Moving outwards from center are the north circumpolar stars. (Assuming a viewing location of approximately 40-50° north latitude).

Surrounding the central area are the seasonal stars. The four seasons (Fall, Win-

44

Figure 4. *Ojibwe Giizhig Anung Masinaaigan*, 'The Ojibwe Sky Star Map.'

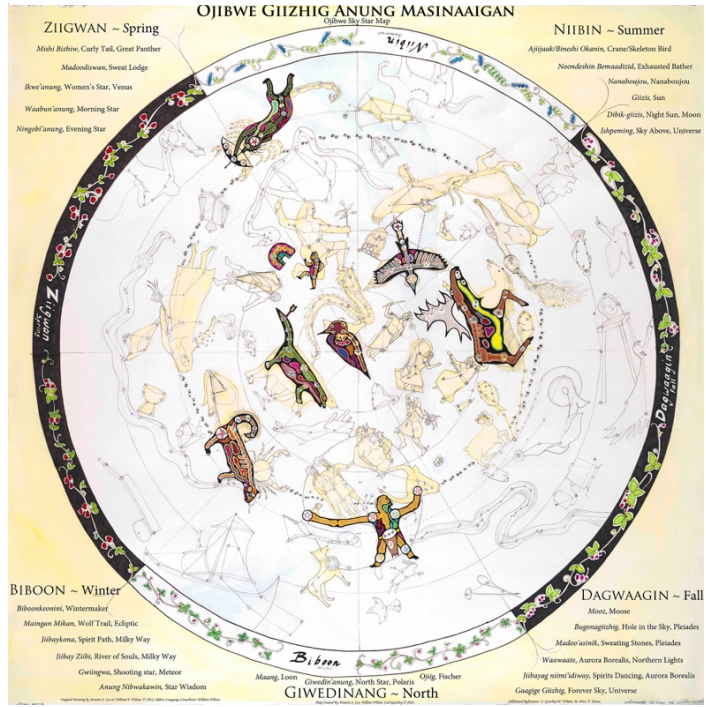

Figure 5. *D(L)akota Makoće Wićaŋhpi Wowapi* 'The Dakota Star Map.'

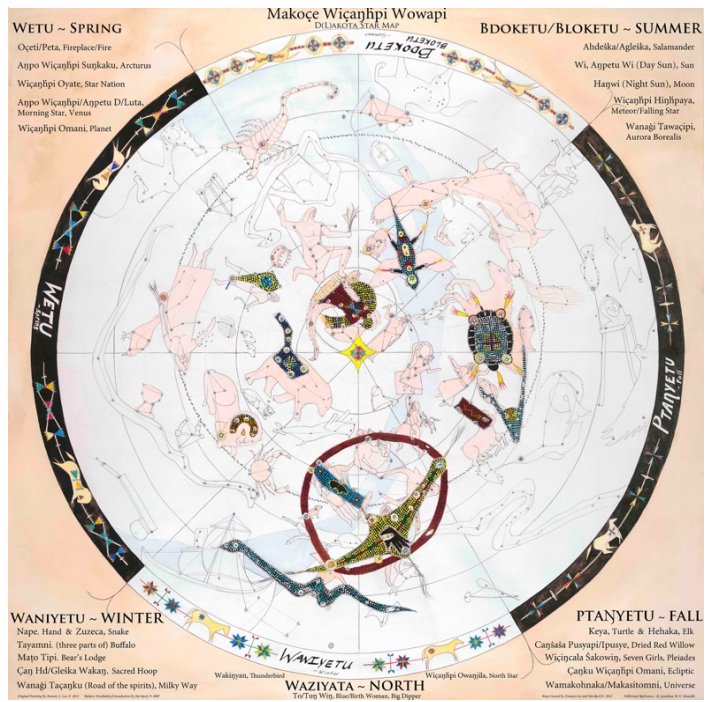

Figure 6. Dakota holder, part of a plansiphere.

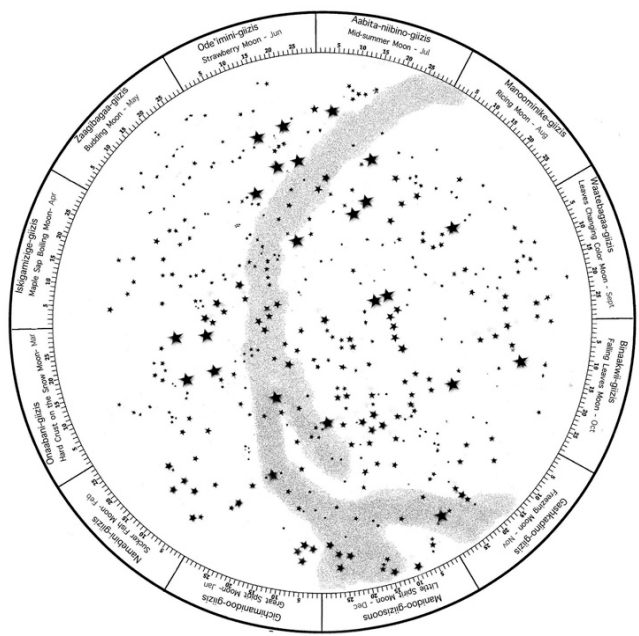

Figure 7. Ojibwe circle.

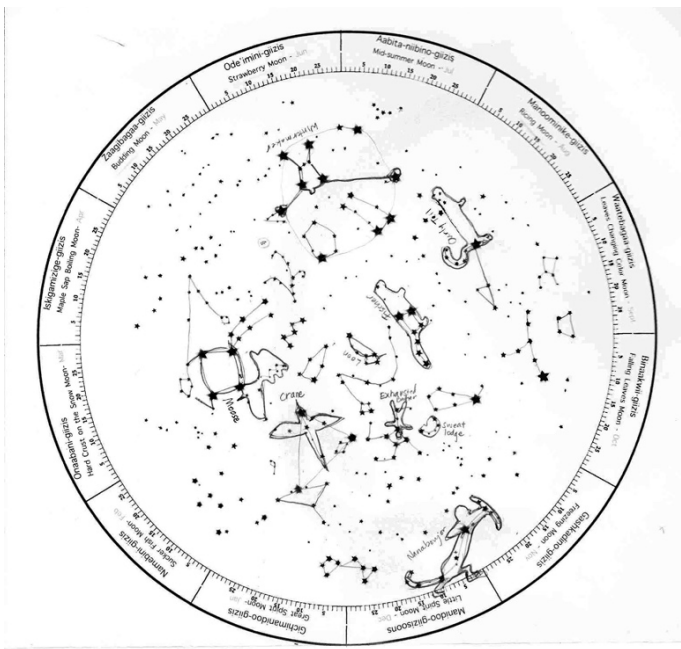

ter, Spring, Summer) can be seen written on the outer border of the map. Each map illustrates the native constellations in brightly colored, clearly marked areas correlated with their seasonal association. Ojibwe constellations were painted in a traditional woodland x-ray style by Wilson. Dakota constellations were painted by the author in a pointillist style to suggest beadwork. In both star maps the Greek constellations are lightly painted in a quiet wash to allow the map-reader some sense of common ground between the Greek and the native Ojibwe and D(L)akota constellations.

*Websites*
Each map has a dedicated webpage. http://web.stcloudstate.edu/aslee/OJIB-WEMAP/home.html, and http://web.stcloudstate.edu/aslee/DA-KOTAMAP/home.html.
These websites contain the following: the star maps (that can be downloaded at no cost); star vocabulary (in Ojibwe/Dakota and English); recorded audio of the star vocabulary in both Ojibwe and Dakota; related curriculum; and upcoming events. The purpose of the website is to allow greater accessibility and communication concerning native star knowledge for local, regional, national and global learners. (See Appendixes) In March 2015, an additional website was created to support *Native Skywatchers* art programming: http://www.nativeskywatchers.org.

*Guidebooks*
In June 2014, two constellation guidebooks were published to accompany each of the star maps: *Ojibwe Sky Star Map Constellation Guide* and *D(L)akota Star Map Constellation Guide.*
    Each booklet contains additional information about individual native constellations, selected teachings, and artwork.

Celestial objects from western European astronomical traditions in the corresponding areas of the night sky are also highlighted throughout the booklet. For example, looking skyward on a fall evening, a person might see: the Ojibwe *Mooz* (Moose), the Dakota *Keya* (Turtle) and the Greek constellation Pegasus, as well as, the Andromeda Galaxy, M31.

*Planispheres*
Tangible resources are essential when working to engage communities in learning the patterns of the night sky. One manipulative that is very useful is called a "planisphere."
    Planispheres are two-dimensional versions of astrolabes that date back to at least the ninth century and perhaps 150 A.D. (Ridpath, 1988). They are star-finding devices consisting of two parts: an inner circle that illustrates a year-round star map and an outer holder. The important idea here is that a person can get a snapshot view of the night sky for any particular time and date, and watch how the sky changes with time. This device is accurate for latitudes 40-50° N and our epoch of Polaris (due to precession there would be some shifting for longer time periods).
    Notice that on the outer perimeter of the star circle (or wheel) are the native Ojibwe and Dakota names for each month. These will vary with local latitudes and longitudes, but are correct for local communities (Fond du Lac and Minneapolis/St. Paul). The months are named for the Moon and significant cultural events for each season. For example: August is "*Manoominike-giizis*" (Ricing Moon) in Ojibwe and March is "*Išta Wicayazaŋ Wi*" (Sore Eyes Moon) in Dakota. (Tables 3 & 4 – Moons) Also significant is a 'blank' star circle that displays individual stars but no constellation lines, names or artwork. This is an excellent way to practice learning the

visual patterns of the constellations in the night sky.

## Educator Collaboration

### Educator Workshop

The native Ojibwe and D(L)akota star maps were designed in large part for the first *Native Skywatchers Educator Workshop* in June 2012 which was co-funded by NASA-MN Space Grant, St. Cloud State University, Fond du Lac Tribal and Community College. A federal science agency, a comprehensive state university, and a two-year tribal community college supported this work. The *Native Skywatchers* team, directed by A. Lee, presented two consecutive days of workshop activities offered at two distinct sites: Fond du Lac Tribal & Community College (FDLTCC) and St. Cloud State University (SCSU) based jointly at the SCSU Planetarium and the American Indian Center (AIC). Participants included: K-12 educators, informal science educators, college educators, planetarians, administrators, and members of the community who are astronomy enthusiasts. The foci of the workshop were to: 1.) give participants information about Ojibwe and D(L)akota star knowledge; 2.) embed star knowledge in an authentic cultural context that includes history, language, art, etc. in addition to astronomy; 3.) create awareness and dialog addressing the protocols and respectful ways of sharing this cultural knowledge. Each year since 2012 the dual workshops have been offered and well attended (~30 participants each workshop). In 2015, an additional workshop was designed and delivered at the Ziibiwing (Ojibwe) Cultural Center in Mt. Pleasant,

Figure 8. Poster for Educator and Community Workshop

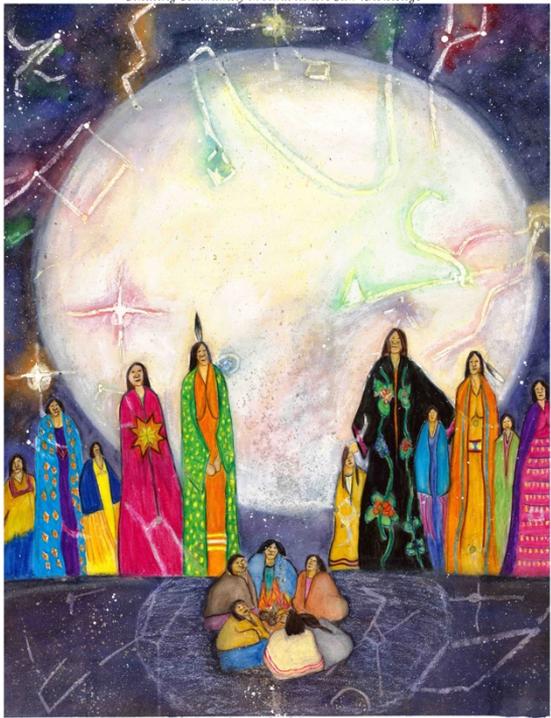

Michigan, co-supported by Central Michigan State University (CMU). The Ziibiwing workshop presented Ojibwe star knowledge and included hands-on art activities related to star knowledge for participants.

### Outreach

The SCSU Planetarium boasts a state-of-the-art fiber optic Chronos star projector that brings more than 8,500 stars and 24 constellations and galaxies to life under a 30-foot dome. Renovated in 2007, at a cost of $790,000, SCSU Planetarium is the only planetarium in the region to offer planetarium shows at no cost. Over 7,000 people visit the SCSU Planetarium each year for shows and astronomy events. *Native Skywatchers* is integrated into the regular

planetarium programming. (A. Lee is also the SCSU Planetarium Director). Educators, scout leaders, educational organizations, etc. can request a planetarium show about Ojibwe and/or D(L)akota star knowledge. In May of 2015 a collaborative effort between *Native Skywatchers*, the SCSU Planetarium, and the St. Cloud School District brought hundreds of middle school students into the SCSU Planetarium for *Native Skywatchers* Planetarium shows.

Curriculum
*Which way is North?*
Annette S. Lee, director of the *Native Skywatchers* research and programming initiative has created curriculum for educators containing Ojibwe and D(L)akota star knowledge that complements the star maps. This includes lesson plans, and worksheets. In many indigenous cultures there is great importance in both ceremony and in everyday life given to the four directions. Here we use this important cultural framework extended to the night sky.

The first exercise is called, *"Which way is North?,"* written by A. Lee (2009). The lesson begins with a discussion around the following four questions:

o  *Which way is North?*
o  *How do you know?*
o  *How do you say North in Ojibwe? In Dakota?*
o  *Why is it important? (both culturally and astronomically)*

This lesson is an important foundation for participants to learn about the patterns of motion in the stars, (assuming northern hemisphere, mid-latitudes); topics include: the north star, the north celestial pole, identifying Ojibwe, D(L)akota, Greek constellations in the north sky, and the larger cultural meaning of these constellations. It is an important part of the discussion to locate the north compass direction in the room at the time of the discussion. This connects the abstract with the 'here and now' and makes the idea tangible. Tape is used to make a capital 'N' on the wall and handwritten signs are made to show the Ojibwe and Dakota words for north: *Giiwedinong* and *Waziyata*. Workshop participants face north, while talking about the North. Indeed our internal and external compasses are aligned and the learning is amplified.

*Follow the Seasons; Follow the Stars*
A second lesson plan written by A. Lee is called, "*Follow the Season; Follow the Stars, Four-Direction Star Gazing*" (2009). This assumes the viewer is facing the south horizon. Again, the meeting room wall is marked with a big letter 'S' using tape and signs indicating the Ojibwe and Dakota words for south: *Zhaawanong* and *Itokagata.* Facing the South, we notice a completely different pattern of motion compared to the apparent stillness of the north night sky. In the south, we have east to west motion. The path of the Sun is highest in the South each day. The Sun transits the local meridian. Here we make a very clear association with the seasons. The basic framework is that whatever season is presently being experienced, that same season of stars will be seen in the night sky a few hours after sunset, facing South. The south direction and the overhead (zenith) are referred to as the 'center stage of the night sky'. Described above is a distinctly indigenous way of teaching that emphasizes learning through direct experience, place-based examples, and relationships. (Cajete, 2000)

**Selected Teachings**
Kapemni

One of the most important underlying ideas in both the Ojibwe and D(L)akota star knowledge, is the idea *"As it is above; it is below."* This is a very old and universal idea found in many indigenous cosmologies. (Campion, 2012) The idea can be visualized by two triangles (or tipis) stacked vertically connected at their apexes.

*Figure 9. Illustration of Kapemni*

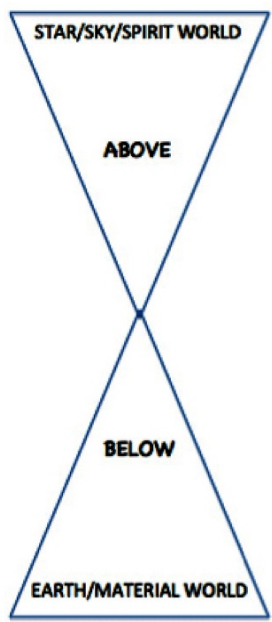

The bottom tipi represents the material, or physical world. The top tipi represents the sky, star, and spirit world. The first part of the teaching is that both realms are real. Both realms are important and meaningful. Furthermore, traditional native people should strive to, either in ceremony or everyday life, acknowledge this relationship. When they do, it is like standing at the apex between the two triangles, which functions like a doorway, and there is a flow between the two worlds. An idea so important, in Dakota there is a single word, *'kapemni'* to describe this teaching.

*"Traditional Lakota believed that ceremonies done by them on earth were also being performed simultaneously in the spirit world. When what is happening in the stellar world is also being done on earth in the same way at the corresponding place at the same time, a hierophany can occur; sacred power can be drawn down…"* (Goodman, 1992)

In D(L)akota many constellations have '*kapemni pairs*.' This means that there is a geographical site located on Earth that is the counterpart to the constellation in the sky. For example, when the Sun is in the constellation *Mato Tipila* (Bear's Lodge, Gemini), this is the traditional time for people to meet at *Pte he Ġi* (Grey Horn Butte, Devil's Tower) and pray. In Ojibwe, the teaching of mirroring the 'above and below' is most often seen in the naming of constellations with animals displaying the same pattern of motion as their celestial counterpart. For example, the Big Dipper is *Ojiig* (Fisher). Known as a cultural hero animal for an important rescue mission, this small weasel-like mammal has many patterns of behavior in both real life and in cultural stories that coincide with the pattern of circumpolar motion as displayed by the Big Dipper (i.e. the seven brightest stars in Ursa Major).

Maang
When looking to the North, in addition to identifying the North Star and the pattern of celestial motion in the north, we can identify the Ojibwe constellation, *Maang* (Loon).

The *Maang* (Loon) constellation encompasses the same set of stars as the Little Dipper or the brightest seven stars in Ursa Minor. The first, most obvious question is: why did the Ojibwe associate this North American aquatic diving bird, the loon, with the most central place in the night sky? There are several answers to this question. The first reason is that the loon

Figure 10. Ojibwe Maang (Loon) constellation.

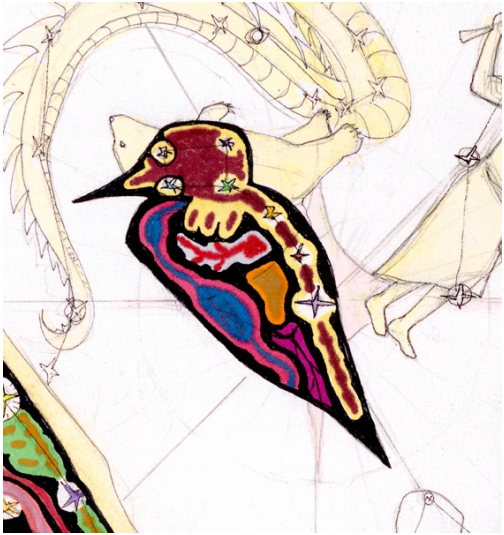

is one of the two leaders in the Ojibwe clan system (a framework of government and organizing society):

*"The Crane and the Loon Clans were given the power of chieftainship. They were given the people with natural qualities and abilities for leadership."* (Benton, 1988)

The North Star—being within one degree of the north celestial pole (NCP)—appears almost motionless as viewed from the ground. All other stars and celestial objects in the entire day and night sky appear to be circling around it. The motionless star is a leader. This leadership is reflected in the star knowledge by naming *Maang*, the Ojibwe clan leader, as the constellation containing the seven brightest stars nearest the motionless point (NCP) or the Little Dipper. *Maang* (Loon) is a leader in the sky, and a leader on the Earth. This illustrates the mirroring of Earth and Sky.

The teachings of the loon are many. One example is that the loon has a very close connection to the water. It avoids going on land, except to nest. It is not uncommon that loons become stranded in parking lots or pasture puddles. Loons need a quarter mile of water to take off. (Evers et. al, 2010) In addition, loons survive physically by diving and spearing fish. Loons need clear, calm water to survive. The sacredness of water for life and survival is found throughout Ojibwe culture. Lastly, the pattern of a black background with many small white dots mirrors a starry night sky. The loon has the stars of the night sky reflected on its back. As described by one elder, when an Ojibwe person hunts the loon, respect is shown by offering tobacco to the loon for giving its life; and the loon is never to be turned upside-down (the backside of the loon must always be facing the sky). (Wilson, 2012) The earth-sky mirroring is respected and maintained even after its death. Ojibwe first language speaker and *Native Skywatchers* team-member, W. Wilson, explains that the word "*maang*" is closely related to the word, "*a maa(ng)*" which means "listen or pay attention." (Wilson, 2012)

## Conclusions

This important work has many branches: interdisciplinary connections in science and culture, formal and informal science education, artwork and art programming, history and heritage, outreach and community wellness. The *Native Skywatchers – Revitalization of Ojibwe and D(L)akota Star Knowledge* research and programming initiative has worked with community members to create meaningful recourses that communicate an ancient and living relationship with the cosmos. It is our goal to build community around native star knowledge. All cultures, throughout human history have had a connection to the stars. (Campion, 2012) It is intended that the *Native Skywatchers* research and programming initiative will help individuals and communities in rebuilding and remembering the native Ojibwe and D(L)akota connection to the stars. Ultimately, it is hoped that this dialog will serve as a stepping-stone to honor and remember all indigenous ways of knowing.

Figure 11. Maang-Doorkeeper of the North", A. Lee, 2014, mixed media on panel.

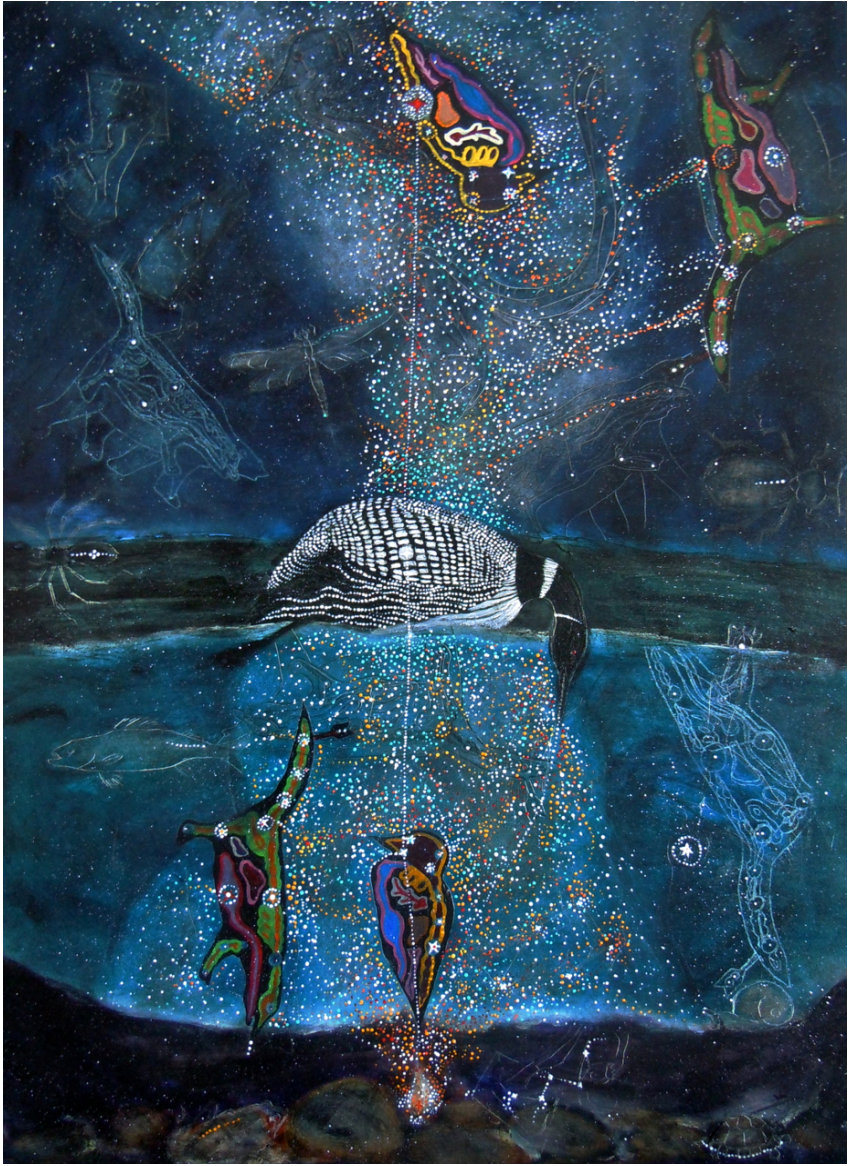

## Acknowledgments

Miigwech. Pidamaya. Thank you.

## Appendix A:
Ojibwe Celestial Terminology

### Table A1. Ojibwe Vocabulary – Seasonal

| Ojibwe | English | Related Greek Constellations |
|---|---|---|
| *Dagwaagin* | Fall | |
| *Mooz* | Moose | Pegasus |
| *Bugonagüzhig* | Hole in the Sky | Pleiades |
| *Madoo'asinik* | Sweating Stones | Pleiades |
| *Biboon* | Winter | |
| *Biboonkeonini* | Winter-maker | Orion, Canis Minor, Taurus |
| *Ziigwan* | Spring | |
| *Mishi bizhiw* | Curly Tail, Great Panther | Leo, Hydra |
| *Madoodiswan* | Sweat Lodge | Corona |
| *Niibin* | Summer | |
| *Ajiijaak/Bine-shi Okanin* | Crane/Skele-ton Bird | Cygnus |
| *Noondeshin Bemaadizid* | Exhausted Bather (Per-son) | Hercules |
| *Nanaboujou* | Nanaboujou | Scorpio |
| *Giwedinang* | North | |
| *Maang* | Loon | Little Dipper |
| *Ojiig* | Fischer | Big Dipper |
| *Giwedin'anung* | North Star | Polaris |

### Table A2. Ojibwe Vocabulary – Objects

| Ojibwe | English |
|---|---|
| *Anung* | Star |
| *Anung aki* | Star World |
| *Dibik-giizis* | Moon ('Night Sun') |
| *Güzis* | Sun |
| *Güzhig* | Sky |
| *Ikwe'anung* | Venus ('Women's Star') |
| *Ningobi'anung* | Venus as Evening Star |
| *Waabun'anung* | Venus as Morning Star |
| *Maingan Mikan* | Ecliptic ('Wolf Trail') |
| *Jiibaykona* | Milky Way ('Spirit Path') |
| *Jiibay Ziibi* | Milky Way ('River of Souls') |
| *Gwiingwa* | Meteor/shooting star |
| *Gaagige Güzhig* | Universe ('Forever Sky') |
| *Ishpeming* | Universe ('The Sky Above') |
| *Waawaate* | Aurora Borealis (North-ern Lights) |
| *Anung Nibwakawin* | Star Knowledge, Wisdom |

### Table A2. (*cont.*)

| Ojibwe | English |
|---|---|
| *Güzhig Anung* | Sky/Star map |
| *Masinaaigan* | |
| *Waabunong* | East |
| *Ningobinong* | West |
| *Giwedinong* | North |
| *Jawanong* | South |

### Table A3. Ojibwe Months/Moon – Fond du Lac Region

| Month | Ojibwe | English |
|---|---|---|
| January | *Gichimanidoo-giizis* | Great Spirit Moon |
| February | *Namebini-giizis* | Sucker Fish Moon |
| March | *Onaabani-giizis* | Hard Crust on the Snow Moon |
| April | *Iskigamizige-giizis* | Maple Sap Boiling Moon |
| May | *Zaagibagaa-giizis* | Budding Moon |
| June | *Ode'imini-giizis* | Strawberry Moon |
| July | *Aabita-niibino-giizis* | Mid-summer Moon |
| August | *Manoominike-giizis* | Ricing Moon |
| September | *Waatebagaa-giizis* | Leaves Chang-ing Color Moon |
| October | *Binaakwii-giizis* | Falling Leaves Moon |
| November | *Gashkadino-giizis* | Freezing Moon |
| December | *Manidoo-giizisoons* | Little Spirit Moon |

# Appendix B
## D(L)akota Celestial Terminology

### Table B1. D(L)akota Vocabulary – Seasonal

| D(L)akota | English | Related Greek Constellations |
|---|---|---|
| *Waniyetu* | Winter | |
| *Nape* | Hand | Orion, Eridanus |
| *Maṫo Tipi/ Maṫo Tipila* | Bear's Lodge | Gemini |
| *Ki Iŋyaŋka Oçaŋku* | Racetrack | Winter Circle |
| *Çaŋ Hd/ Gleška Wakaŋ* | Sacred hoop | Winter Circle |
| *Inipi/ Initipi* | Sweat lodge | Winter Circle |
| *Tayamni* | Buffalo in three parts | Orion, Canis Major, Pleiades |
| *Tayamni pa* | Buffalo embryo head | Pleiades |
| *Tayamni cutuhu* | Buffalo embryo ribs | Betelgeuse & Rigel |
| *Tayamni caŋkahu* | Buffalo embryo backbone | Orion's belt |
| *Tayamni siŋte* | Buffalo embryo tail | Sirius |
| *Zuzeca/ Zuzuheča* | Snake | Columbia, Puppis, Canis Major |
| *Ptaŋyetu* | Fall | |
| *Keya* | Turtle | Pegasus |
| *Caŋšaša Pusyapi/ Ipusye* | Dried Red Willow | Aries, Triangulum |
| *Heḣaka/ Upaŋ* | Elk | Pisces |
| *Wiċiŋyaŋna Šakowiŋ/ Wiċiŋcala Šakowiŋ* | Seven Girls | Pleiades |
| *Bdoketu/ Bloketu* | Summer | |
| *Ahdeška/ Agleška* | Salamander | Cygnus |
| *Wetu* | Spring | |
| *Oċeti/Peta* | Fireplace/ Fire | Leo |

### Table B1. (*cont.*)

| | | |
|---|---|---|
| *Itkob u* | Arcturus ('going toward') | Bright star in Bootes |
| *Ihuku Kiĝle* | Arcturus ('underwent it') | Bright star in Bootes |
| *Aŋpo Wiċaŋḣpi Suŋkaku* | Arcturus ('younger brother of Morning star') | Bright star in Bootes |
| *Waziyata* | North | |
| *To Wiŋ/ Tuŋ Wiŋ* | Blue Woman/ Birth Woman | Big Dipper - inside Bowl |
| *Wiċakiyuhapi* | Stretcher | Big Dipper - Bowl |
| *Wašiḣdapi/ Wašiglapi* | Mourners | Big Dipper - Handle |
| *Maŋka/Maka* | Skunk | Big Dipper |
| *Wiċakiyuhapi/ Can cinkska* | Dipper/ Wooden Spoon | Big Dipper |
| *Oċeti Šakowiŋ* | Seven sacred rites/Council fires | Big Dipper |
| *Wiċaŋḣpi wa-ziyata/Wiċaŋḣpi Owaŋjila* | North Star ('Star that stands in one place') | Polaris |
| *Wakiŋyaŋ* | Thunderbird | Draco, Ursa Minor |

### Table B2. D(L)akota Vocabulary – Objects

| D(L)akota | Celestial Object |
|---|---|
| *Wiċaŋḣpi* | Star |
| *Wiċaŋḣpi Oyate* | Star Nation |
| *Haŋhepi Wi / Haŋyetu Wi/Haŋwi* | Moon ('Night Sun') |
| *Anog Ite* | Moon ('Double Faced Woman') |
| *Wi, Aŋpetu Wi* | Sun ('Day Sun') |
| *Aŋpo Wiċaŋḣpi / Aŋpetu D/Luta* | Venus - Morning Star |
| *Çaŋku Wiċaŋḣpi Omani/ Maḣpiya Maka Iciyagle* | Ecliptic |
| *Wanaĝi Taċaŋku (Road of the spirits/Ghost trail)* | Milky Way |
| *Wiċaŋḣpi Hiŋhpaya/ Wiaḣpiḣinḣpaya/ Woḣpe Wakaŋ* | Meteor/Falling star |
| *Wamakoḣnaka/ Wamakḣog-naka/ Makasitomni* | Universe |

Table B2. (cont.)

| | |
|---|---|
| *Wanaǧi Tawaċipi (Spirit Dancers)/Maḣpiyatayiŋ/ Wiyosaya* | Aurora Borealis (Northern Lights) |
| *Wiċaŋḣpi Siŋtetuŋ/ Wiċaŋḣpi Siŋte Yukan/ Wicaŋpisiŋtetoŋ* | Comet |
| *Makoċe Wiċaŋḣpi Wowapi* | Star map |
| *Wiċaŋḣpi Omani/Wiċaŋḣpi Nuni/Wiċanḣpi Sa* | Planet |
| *Wiaceic′ iti* | Sundogs ('Sun making fire') |
| *Witha Wit′ e* | Solar Eclipse ('Sun dies') |
| *Haŋwitha* | Lunar Eclipse |
| *Wiċaŋḣpi Tiospaye* | Constellations ('Extended family') |
| *Okakśe Taŋka Wiċaŋḣpi Ota/Wiċaŋḣpi Optaye Taŋka* | Galaxies |
| *Wiċaŋḣpi Oyate* | Groups of galaxies ('Nation') |
| *Bdoke cokaya/Bloke cokaya/ Aŋpawi* | Summer solstice ('Morning Sun') |
| *Waniyetu cokaya/Nahomni* | Winter solstice ('Swing around') |
| *Wetu Aŋpa Haŋyetu Iyehaŋtu* | Spring Equinox |
| *Ptaŋyetu Aŋpa Haŋyetu Iyehaŋtu* | Fall Equinox |
| *Omaka/Makoncage* | Seasons ('Earth grows with time change') |
| *Waziyata* | North |
| *Itokagata* | South |
| *Wiohinyanpata* | East |
| *Wiyoḣpeyata* | West |
| *Wankantu/Waŋkatika* | Above |
| *Kutakiya/Kutkiya* | Below |
| *Ċokata/Ċokaya* | Center |
| *Tayamni pa* | Buffalo embryo head (Pleiades) |
| *Tayamni cutuhu* | Buffalo embryo ribs (Betelgeuse and Rigel) |
| *Tayamni caŋkahu* | Buffalo embryo backbone (Orion's belt) |
| *Tayamni siŋte* | Buffalo embryo tail (Sirius) |

Table B3. D(L)akota Months/Moon – Minneapolis Region

| Month | D(L)akota | English |
|---|---|---|
| January | *Witehi Wi/ Wiot′ ehika Wi/ aŋkapopa Wi* | Hard/Difficult Moon/Tree Popping Moon |
| February | *Wi aṫ′a Wi / Aŋpetu Numnuŋpa Wi/ Wicata Wi/ Cannapopa Wi* | Raccoon Moon/ Moon when many die/ Two different kinds of days Moon/ Moon of Popping Trees |
| March | *Iṡta Wicayazaŋ Wi* | Sore eyes Moon |
| April | *Watopapi Wi/ Maǧa Okada Wi/ Wokada Wi/ Wihakakta Wi* | Moon when streams are open/Goose egg-laying Moon/Egg laying Moon/ Moon of Fattening |
| May | *Woźupi Wi* | Planting Moon |
| June | *Waźuśtecaśa Wi/Wipazuka Waste Wi* | Strawberry Ripening Moon/Moon of Good Berries Moon |
| July | *aŋpasapa Wi/ aŋpaśa Wi/ Waśuŋpa Wi* | Moon when the Chokecherries are Ripe/ Moon when the geese shed their feathers |
| August | *Wasutuŋ Wi* | Harvest Moon |
| September | *Psiŋhnaketu Wi/ Taśaheca Hakikta Wi/ Wayuksapi Wi/ Canwapegi Wi* | Moon when the rice is laid up to dry/ Moon when the chipmunk looks back/Corn harvesting Moon/Moon of Brown Leaves |
| October | *aŋwaḣpekasna Wi/ Wi Waźupi* | Trees shaking off the leaves Moon/Drying rice Moon |
| November | *Takiyuḣa Wi/ Waniyetu Wi* | Deer Rutting Moon/Moon of Rutting Deer |
| December | *Tahecapśuŋ Wi* | Deer Antler Shedding Moon |